\documentclass{article}

\RequirePackage{fix-cm}

\newcommand{\ket}[1]{\left| #1 \right>} 
\newcommand{\bra}[1]{\left< #1 \right|} 
\let\baraccent=\= 
\renewcommand{\=}[1]{\stackrel{#1}{=}} 

\usepackage{graphicx,epstopdf}
\usepackage{dcolumn}
\usepackage{bm}

\begin{document}           

%
%

\title{Ground state energy density of the quantum harmonic oscillator}


\author{Firmin J. Oliveira \\
P. O. Box 10882 \\ Hilo, Hawaii, U.S.A.  96721-5882   \\ Email address: firmjay@hotmail.com \\ } 


\date{Received: date / Accepted: date}

\maketitle

\begin{abstract} 

The total energy of the ground state of the quantum harmonic oscillator is obtained with minimal assumptions.
The vacuum energy density of the universe is derived and a cutoff  frequency is obtained
for the upper bound of the quantum harmonic oscillator. 

\end{abstract}


\section{Introduction}

In this paper we find an expression for the ground state energy of the quantum harmonic oscillator (QHO) to obtain the 
vacuum energy density of the universe. We use  the addition of velocities of 
Einstein's Special Relativity theory to derive a contraction of the Planck time.
Using Heisenberg's Uncertainty Principle we obtain an expression for the total energy of the QHO ground state.
We also obtain a cutoff frequency for the QHO.
This paper is a parallel version of an earlier e-publication by the author\cite{oliveira-1} which was based on 
Carmeli's Cosmological Special Relativity theory\cite{carmeli-1}.

\section{Addition of velocities and the contracted  Planck time}

Consider an inertial reference frame  $K_1$  which moves at a  velocity $v_1$ along  the $x$ axis of inertial frame $K$.  
If  an object is measured to have a velocity $v_2$ along the $x_1$ axis relative to the origin of frame $K_1$ 
then the velocity $v_{1+2}$ of the object relative to the origin of frame 
$K$ is given by the addition of velocities of Einstein's Special Relativity theory\cite{einstein-1},
\begin{equation}
v_{1+2} = \frac{ v_1 + v_2} { 1 + v_1 v_2 / c^2},    \label{eq:v1+2}
\end{equation} 
where $c$ is the speed of light in vacuum.  Assume that $v_1 = v$, $v_2 = \delta{v'}$ and $v_{1+2} = v + \delta{v}$,
where $\delta{v'} \ll c$.  Substituting these variables into (\ref{eq:v1+2}) gives
\begin{equation}
v + \delta{v}  =  \frac{ v + \delta{v'} } { 1 + v \, \delta{v'} / c^2 }
  \approx \left(v + \delta{v'} \right) \left(  1  - \frac{v \, \delta{v'}} { c^2 } \right) 
  \approx v + \left( 1 - \frac{v^2} {c^2} \right) \delta{v'}, \label{eq:v+del_v}
\end{equation}
where approximations are made due to $\delta{v'} \ll c$.  Simplifying  (\ref{eq:v+del_v}) yields
\begin{equation}
\delta{v}  \approx  \left( 1 - \frac{v^2} {c^2} \right) \delta{v'}.  \label{eq:delta_v}
\end{equation}
Assume velocity $v$ is very close to the speed of light $c$ such that it is given by
\begin{equation}
v = c - \delta{v'}.  \label{eq:v_def}
\end{equation}
For this value of $v$ the second term in the first factor on the right hand side of (\ref{eq:delta_v}) is given by
\begin{equation}
-\frac{v^2} {c^2} =  -\frac{\left(c - \delta{v'} \right)^2} {c^2}  \approx -1 + \frac{2 \delta{v'} } {c},  \label{eq:-v2/c2}
\end{equation}
where we make the approximation again since $\delta{v'} \ll c$.
Substituting (\ref{eq:-v2/c2}) into (\ref{eq:delta_v}) and simplifying we obtain
\begin{equation}
\delta{v}  \approx  \frac{2 \left(  \delta{v'} \right)^2} {c}.  \label{eq:delta_v_approx}
\end{equation}

Now, define the acceleration constant $a_0 = c H_0$, where $H_0$ is Hubble's constant, and
use it to define the expansion velocity parameter
\begin{equation}
{\cal{V}} = a_0 {\cal{T}}_P,  \label{eq:V_def}
\end{equation}
where the Planck time\cite{planck-1} ${\cal{T}}_P = \sqrt{\hbar G / c^5}$,  where $\hbar$ is the 
reduced Planck constant and $G$ is Newton's constant.  Eq. (\ref{eq:V_def}) is just Hubble's law\cite{hubble-0}
given by ${\cal{V}} = H_0 {\cal{L}}_P$ where ${\cal{L}}_P = c {\cal{T}}_P$ is the Planck length.
It is a fundamental principle of Special Relativity that the laws of physics are invariant for all observers in inertial 
reference frames, from which it follows that the physical constants are the same for all inertial observers. 
Since $\hbar$, $G$, $c$ and $H_0$ are physical constants,
hence $a_0$ and therefore ${\cal{V}}$ are both constants  in any inertial reference frame. 
Then, substituting from (\ref{eq:V_def}) for the  velocity  in  (\ref{eq:delta_v_approx})  
such that $\delta{v'} = {\cal{V}} = \sqrt{\hbar G / c^3 } H_0$, we obtain the expression for 
the contracted expansion velocity 
\begin{equation}
{\cal{V}}_C  = \delta{v}  = \frac{2 {\cal{V}}^2} {c}  =  \frac{2 \hbar G H^2_0} {c^4}.  \label{eq:V_C_def}
\end{equation}
What this means is that while the observer in $K_1$ detects that the object has an expansion velocity ${\cal{V}}$,  
the observer in $K$ will say that the object actually has an expansion velocity of
 ${\cal{V}}_C$ relative to the origin of  $K_1$
so that the expansion velocity of the object in $K$ is $ v_{2K} = c -  {\cal{V}} + {\cal{V}}_C$.
In order to obtain a time value from the contracted velocity, divide $a_0$ into (\ref{eq:V_C_def}) to get the time interval
\begin{equation}
{\cal{T}}_{PC}  =  \frac{{\cal{V}}_C} {a_0}  =  \frac{2 \hbar G H_0}{c^5}.  \label{eq:T_PC}
\end{equation}
We call ${\cal{T}}_{PC}$ the contracted Planck time since it  has the form 
\begin{equation}
 {\cal{T}}_{PC} = 2 \left({\cal{T}}_P \right)^2 H_0.  \label{eq:T_P0_T_P}
\end{equation}

With ${\cal{T}}_P \approx 5.39 \times 10^{-44} {\rm s}$ 
and $H_0 \approx 71.9 \, {\rm km  \, s^{-1}  \, Mpc^{-1} }\approx 2.33 \times 10^{-18} {\rm s}^{-1}$\cite{hubble-1},    
we get ${\cal{T}}_{PC} \approx 1.35 \times 10^{-104} {\rm s}$.  The expansion velocity parameter has a value
${\cal{V}} \approx 3.77 \times 10^{-51} \, {\rm cm \, s}^{-1}$ while the contracted expansion velocity has a value
${\cal{V}}_C \approx 9.46 \times 10^{-112} \, {\rm cm \, s}^{-1}$. The value of the acceleration constant is
$a_0 \approx 6.99 \times 10^{-8} \, {\rm cm \, s^{-2}}$.

\section{Quantum harmonic oscillator ground state energy}

For a simple linear quantum harmonic oscillator (QHO) in one dimension the energy levels for the plane wave modes ${\bf k}$
 and frequencies $\omega_{{\bf k}}$ are given by\cite{rugh-1,marsh-1}
\begin{equation}
E\left( {\bf k} \right)_n  =  \left( n + \frac{1} {2} \right)  \hbar \omega_{{\bf k}},  \label{eq:E_QHO}
\end{equation}
where the quantum number $n = 0, 1, 2, ...$ is the state number.  The ground state energy for the frequency
$\omega_{{\bf k}}$ is given for $n = 0$, 
 \begin{equation}
E\left( {\bf k} \right)_0  =   \frac{1} {2}  \hbar \omega_{{\bf k}}.  \label{eq:E_QHO_0}
\end{equation}
For three linear QHO's oriented along orthogonal axes, and for two polarizations, the ground state energy of the 
electromagnetic field is given by
\begin{equation}
E\left( {\bf k} \right)_0  =  2 \left( \frac{3} {2} \right)  \hbar \omega_{{\bf k}}  
       =3 \, \hbar \omega_{{\bf k}}.  \label{eq:E_QHO_3d0}
\end{equation}
To get the total energy density $\rho_{vac}$ for all  modes ${\bf k}$ in the ground state $\ket{0}$ 
we must sum over all oscillator mode frequencies between zero and a finite cut-off frequency $\omega_{max}$
to obtain the expected value
\begin{equation}
\rho_{vac}   =  \bra{0} \hat{\rho} \ket{0}  =  \frac{E} { V}  
                = \frac{1} { V} \Sigma_{{\bf k}} \left(  3 \, \hbar \omega_{{\bf k}} \right)
                \approx   \frac{ 1 } {8  \pi^3 c^3 } 
                                  \int^{\omega_{max}}_0 {3 \hbar \omega \left( 4 \pi \omega^2 \right)  d\omega} ,
                    \label{eq:rho_0_1} 
\end{equation}
where $E$ is the total energy of the ground state and $V$ is the volume of the universe.
Performing the integration in  (\ref{eq:rho_0_1}), the relationship between $\rho_{vac}$ and $\omega_{max}$
is given by
\begin{equation}
\rho_{vac} =  \left( \frac{3 \hbar \omega^4_{max}} { 8 \pi^2 c^3}  \right).  
       \label{eq:rhovac_and_wmax}
\end{equation}
We can obtain an expression for the density by   defining the energy $E$ in terms of the contracted Planck time ${\cal{T}}_{PC}$ according to  Heisenberg's 
Uncertanty Principle\cite{heisenberg-1} of the form 
\begin{equation}
\Delta{E} \, \Delta{t}    \ge  \hbar.   \label{eq:HUP}
\end{equation}
Taking $\Delta{t} = {\cal{T}}_{PC}$ from (\ref{eq:T_PC}) and using (\ref{eq:HUP})  with $E = \Delta{E}$,
we assume for the ground state the total energy
\begin{equation}
E = \frac{\hbar} { {\cal{T}}_{PC} }.   \label{eq:E_total}
\end{equation}
For a universe with Hubble radius $R_H = c / H_0$ the volume is given by
\begin{equation}
V = \frac{ 4 \pi } {3 } \left(\frac{c} {H_0} \right)^{3}.  \label{eq:Volume}
\end{equation}
Substituting  for $E$ and $V$ from  (\ref{eq:E_total}) and (\ref{eq:Volume})  into (\ref{eq:rho_0_1})
we get the vacuum energy density\cite{fahr-1}
\begin{equation}
\rho_{vac} = \frac{E} {V} = \left(  \frac{\hbar} { {\cal{T}}_{PC} } \right)  \left ( \frac{3 H^3_0} {4 \pi c^3} \right)
                   =   \frac{ 3 c^2 H^2_0 } { 8 \pi G},    \label{eq:rho_vac_result}
\end{equation}
where the last right hand side expression is identical in form to the critical energy density $\rho_c$ 
of standard cosmology. We note that this expression for  $\rho_{vac}$ was obtained through
a general application of the Hubble law, Special Relativity theory and  the Heisenberg Uncertainty Principle.
Solving (\ref{eq:rhovac_and_wmax}) for $\omega_{max}$ using (\ref{eq:rho_vac_result}) we obtain
\begin{equation}
\omega_{max}  =  \left( \frac{\pi  H^2_0} { {\cal{T}}^2_P } \right)^{1/4}.  \label{eq:omega_max}
\end{equation}

Substituting values for the parameters, the value of the cut-off frequency (\ref{eq:omega_max}) is given by
\begin{equation}
\omega_{max}  =\left( \frac{\pi c^5 H^2_0} {\hbar G}  \right)^{1/4} \approx  8.75 \times 10^{12} \, {\rm Hz}.
      \label{eq:omega_max_value}
\end{equation}
The oscillator ground state energy at the cutoff frequency is
\begin{equation}
\epsilon_{max}   =  3 \hbar \omega_{max}  \approx 0.0173 \, {\rm eV},  \label{eq:eps}
\end{equation}
which is $0.067 \%$ of the upper limit for the electron neutrino rest mass energy of $26  \,\, {\rm eV}$ 
of the {\em  Kamiokande II} experiment\cite{sato-1},
also just $0.69 \%$ of the upper limit for the electron antineutrino rest mass energy of $2.5 \,\, {\rm eV}$
of the {\em Troitsk neutrino mass} experiment\cite{troitsk-1} and
between $4.3 \%$ to  $8.6 \%$ of the upper limit of the summed neutrino rest mass energies
$\sum{m_{\nu} c^2} <( 0.2 - 0.4 )  \, {\rm eV}$ from\cite{goobar-1} and the {\em Planck} experiment\cite{ade-1}.
The value of the vacuum energy density $\rho_{vac}$ $\approx 8.73 \times 10^{-9}$ ${\rm erg \, cm^{-3}}$
$\approx 9.71 \times 10^{-30}  {\rm gm \, c^2 \, cm^{-3} }$
or equivalent to about $5.8$ Hydrogen (HI) atoms per cubic meter. 
The cosmological constant\cite{fahr-1,weinberg-1,maia-1} is expressed by
\begin{equation}
\Lambda  =  \kappa \rho_{vac}, \label{eq:Lambda}
\end{equation}
where $\kappa = 8 \pi G / c^4$ is Einstein's constant.
Substituting for the parameters we get a value $\Lambda \approx  1.81 \times 10^{-56} \, {\rm cm^{-2} }$,
for a flat universe with no matter.   This compares with the  {\em High-Z Supernova Search Team} experiment\cite{riess-1} 
where $\Lambda \approx  1.07 \times 10^{-56} \, {\rm cm^{-2} }$  for a flat universe with 
$H_0 \approx 65.2 \, {\rm km \, s^{-1} \, Mpc^{-1}}$, 
with vacuum density parameter $\Omega_{\Lambda} \approx 0.72$  
and mass density parameter $\Omega_M \approx 0.28$.

\section{Conclusion}

From a general application of the Hubble law and Special Relativity we obtained a time interval ${\cal{T}}_{PC}$
which  is $60$ orders of magnitude smaller than the Planck time ${\cal{T}}_P$.  By the Heisenberg
Uncertainty Principle we associated this time with energy and derived the
vacuum energy density which, from  (\ref{eq:rhovac_and_wmax})  and  (\ref{eq:omega_max}) , 
can be put in the form
\begin{equation}
 \rho_{vac}   =  \frac{3 \hbar} { 8 \pi c^3} \frac{ H^2_0 } { {\cal{T}}^2_P }.
    \label{eq:rho_vac_last}
\end{equation}
Since the ground state energy cutoff (\ref{eq:eps})  is approximately  $4 \%$ to $8 \%$ of the present day upper limits 
of the sum of neutrino rest mass energies, and as the neutrino mass upper limits appear to be diminishing by an order of
magnitude with more sensitive experiments, it is probably not too far-fetched to speculate\cite{sidharth-1,sidharth-2}
that  $\epsilon_{max} \approx  \sum{m_{\nu} c^2}$.


\begin{thebibliography}{}

  \bibitem{oliveira-1} Oliveira, F. J.: First instance of cosmic time and the vacuum density.
    arXiv:gen-ph/1608.06525v1

 \bibitem{carmeli-1} Carmeli, M.:
   Cosmological Special Relativity, 2nd edn.
   World Scientific,  Singapore (2002)

  \bibitem{einstein-1} Einstein, A.: Ann. Phys. {\bf 17}, 891 (1905); English translation in: The principle of relativity.
    (Dover, New York, 1923), pg. 35.

  \bibitem{planck-1} Planck, M.: S.-B. Preuss. Akad. Wiss.,  S.479-480 (1899)

  \bibitem{hubble-0} Hubble, E.:
    A Relation between Distance and Radial Velocity among Extra-Galactic Nebulae.
    Proc. Natl. Acad. Sci. U.S.A.  {\bf 15}(3) 168-173 (1929).
    doi:10.1073/pnas.15.3.168

  \bibitem{hubble-1}
    Bonvin, V.,  Courbin, F.,  Suyu, S. H., Marshall, P. J., Rusu, C. E., Sluse, D., Tewes, M., Wong, K. C., Collett, T.,
    Fassnacht, C. D., Treu, T., Auger, M. W., Hilbert, S., Koopmans, L. V. E., Meylan, G., Rumbaugh, N., Sonnenfeld, A.,
   and Spiniello, C.:
    H0LiCOW – V. new COSMOGRAIL time delays of HE 0435-1223: $ H_0$ to 3.8\% precision from strong lensing in a flat $\Lambda CDM$ model.
   MNRAS {\bf 465}(4) 4914–4930 (2016).
  doi:10.1093/mnras/stw3006.
  arXiv:1607.01790


  \bibitem{rugh-1} Rugh, S. E. and Zinkernagel, H.:      
   The qantum vacuum and the cosmological
    constant problem. Studies in History and Philosophy of Science Part B:
    Studies in History and Philosophy of Modern Physics {\bf 33}(4) 663-705 (2002).
    doi:10.1016/S1355-2198(02)00033-3. arXiv:hep-th/0012253

 \bibitem{marsh-1} Marsh, G. E.:
    The vacuum and the cosmological constant problem.
    arXiv:0711.0220

  \bibitem{heisenberg-1} Heisenberg, W.: {\"{U}}ber den anschaulichen Inhalt der quantentheoretischen Kinematik und Mechanik.  
    Zeitschrift für Physik.  {\bf 43} (3–4) 172–198 (1927).
    doi:10.1007/BF01397280

  \bibitem{sato-1} Sato, K., Suzuki, H.:    
    Analysis of neutrino burst from the supernova 1987A in the Large Magellanic Cloud.
    Phys. Rev. Lett. {\bf 58}(25)  2722 (1967).
    doi:10.1103/PhysRevLett.58.2722

  \bibitem{troitsk-1} Lobashev, V. M., Aseev, V. N., Belesev, A. I., Berlev, A. I., Geraskin, E. V.,
    Golubev, A. A.,  Kazachenko, O, V., 
     Kuznetsov, Yu. E., Ostroumov, R. P.,  Rivkis, L. A.,  Stern,  B. E., Titov, N. A.,  Zadoroghny, C. V., 
     Zakharov and Yu. I.:
    Direct search for neutrino mass and anomaly in the tritium beta-spectrum: Status of ``Troitsk neutrino mass" experiment.
    Nuclear Phys. B  (Proc. Suppl.)
    {\bf 91} (1–3)  280-286 (2001).
    doi:10.1016/S0920-5632(00)00952-X

  \bibitem{goobar-1} Goobar, A., Hannestad, S., M{\"{o}}rtsell, E., Tu, H.: 
   The neutrino mass bound from WMAP 3 year data, the baryon acoustic peak, the SNLS supernovae and the
  Lyman-$\alpha$ forest.
   J. Cosmol. Astropart. Phys. {\bf 6}  019  (2006).   
   doi:10.1088/1475-7516/2006/06/019.    
   arXiv:astro-ph/0602155

 \bibitem{ade-1} Planck Collaboration: Ade, P. A. R.,  Aghanim, N., Arnaud, M., et al.:
    Planck 2015 results XIII. Cosmological parameters.
    Astron. Astrophys. {\bf 594} A13 (2016).
    doi:10.1051/0004-6361/201525830.
    arXiv:astro-ph/1502.01589



\bibitem{fahr-1} Fahr, H. J.:
    Cosmological Consequences of Scale-Related Comoving Masses for Cosmic Pressure, Mass, and Vacuum Energy Density.
    Found. Phys. Lett. {\bf 19}(5) 423-440 (2006).
    doi:10.1007/s10702-006-0902-z


  \bibitem{weinberg-1} Weinberg, S.:
    The cosmological constant problem.
    Rev. Mod. Phys. {\bf 61}(1) 1-23 (1989).
    doi:10.1103/RevModPhys.61.1



  \bibitem{maia-1} Maia, M. D., Capistrano, A. J. S. and  Monte, E. M.:
     The nature of the cosmological constant problem.
     Int. J. Mod. Phys. A {\bf 24} 1545 (2009).
     doi:10.1142/S0217751X09044978.
     arXiv:gr-qc/0905.3655

 

  \bibitem{riess-1}  Riess, A. G., Filippenko, A. V., Challis, P., Clocchiatti, A., Diercks, A., Garnavich, P. M., Gilliland, R. L.,
     Hogan, C. J.,  Jha, S., Kirshner, R. P., Leibundgut, B., Phillips, M. M., Reiss,  D., Schmidt, B. P., Schommer, R. A., 
    Smith, R. C., Spyromilio, J.,  Stubbs, C., Suntzeff, N. B. and Tonry, J.:
    Observational evidence from supernovae for an accelerating universe and a cosmological constant.
    Astron. J. {\bf 116}(3) 1009–1038 (1998).
    doi:10.1086/300499.
    arXiv:astro-ph/9805201

  \bibitem{sidharth-1} Sidharth, B. G.:
    A note on the cosmic neutrino background and the cosmological constant.
    Found. Phys. Lett. {\bf 19}(7) 757-759 (2006).
    doi:10.1007/s10702-006-1063-9

  \bibitem{sidharth-2} Sidharth, B. G.:
    A model for neutrinos.
    Int. J. Theor. Phys. {\bf 52}(12) 4412-4415 (2013).
    doi:10.1007/s10773-013-1759-0.
    arXiv:gen-ph/0904.3639



    

\end{thebibliography}
\end{document}